\documentclass[12pt]{iopart}
\usepackage{graphicx}
\begin{document}

\title{Gibbs paradox and a possible mechanism of like-charge attraction in colloids}

\author{Chi-Lun Lee}
\ead{chilun@ncu.edu.tw}
\address{Department of Physics, National Central University, Jhongli 32001, Taiwan}

\author{Yiing-Rei Chen}
\address{Department of Physics, National Taiwan Normal University, Taipei 11677, Taiwan}

\begin{abstract}
Based on a reconsideration of the Gibbs paradox, we show that a residual, non-extensive term in entropy turns up upon mixing identical particles, whether they are indistinguishable or not.  The positive contribution from this residual entropy leads to a decrease in free energy, and we suggest that this entropic mechanism may serve as a source of like-charge attractions between a pair of colloidal particles or other macroions.  For a system of two colloidal particles along with their neutralizing counterions, such decrease in free energy is of a few thermal energies and therefore crucial to the effective interaction between the particles.
\end{abstract}

\maketitle

\section{Introduction}
The phenomena of the so-called ``like-charge attraction'' among colloidal particles have been observed since the last decade\cite{Fraden94, Tinoco96, Grier96, Peng05, Grier97, Tata08, Pollack09}.  It was first observed in a solution that is confined between glass walls with a narrow separation\cite{Fraden94, Tinoco96, Grier96} as well as metastable colloidal crystals\cite{Grier97}, as quite recently like-charge attraction was also observed in bulk colloidal solutions\cite{Tata08, Pollack09}.  The range of attraction is comparable to the colloidal size, which is in the order of one micrometer.  This relatively long-range attraction among colloidal particles, which cannot be explained through the use of van der Waals force, is generally believed to result from electrostatic origin. Meanwhile, similar effects of like-charge attraction have been observed in polyelectrolyte solutions\cite{Schmitz84,Schmitz93, Wong03} and other macroionic systems\cite{Liu04, Liu05} as well.  While no definite answer is given, mechanisms of the attractive force among these mesoscopic objects have been proposed by the Sogami-Ise theory\cite{Sogami84, Ise10}, the collective fluctuations of counterions\cite{Liu97, Wong03, Lee10}, possible charge inversions in colloids\cite{Tinoco98}, and even hydrodynamic effects of the confining glass plates\cite{Squires00}.

In this work we introduce a possible mechanism to account for the like-charge attraction based on entropic considerations.  To do this, necessarily one needs to go through, with caution, the statistical definition of entropy, as well as the debate it has introduced in the past.  We shall start our discussion with a review on the well-known Gibbs paradox\cite{Gibbs, Jaynes92, Swendsen06}.

\section{Gibbs paradox revisited}
\begin{figure}
  \includegraphics[width = 7cm]{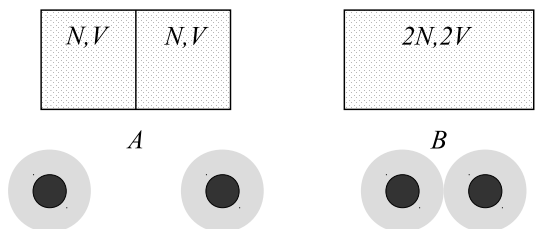}
  \caption{Illustration of the Gibbs paradox and its analogy in a system of two colloidal particles. Case $A$ refers to a unmixed bisected state, while case $B$ refers to the state where the bisecting partition is removed, or the mixed state where counterion clouds come close.}
  \label{Fig1}
\end{figure}

First let us review the Gibbs paradox with a simple lattice model.
Consider $N$ identical particles distributed among $V$ lattice sites in a container.
We assume no interaction except volume exclusion, such that each site can be occupied by at most one particle.
According to the Boltzmann entropy formula one has
\begin{equation}
S_0 = k_{\rm B} \ln \frac{V!}{(V-N)!} \approx k_{\rm B} [ V\ln V - N - (V-N)\ln (V-N)]
\label{Boltzmann}
\end{equation}
with the use of Stirling's approximation $V! \approx V\ln V - V$ and so on.  If one joins two such containers together (as illustrated in Fig.~\ref{Fig1}), the total entropy of the combined system before the removal of the partition is
\begin{equation}
S_A =2S_0 \approx 2k_{\rm B} [ V\ln V - N - (V-N) \ln (V-N)] \, .
\label{S_A}
\end{equation}
After removing the partition the total Boltzmann entropy becomes
\begin{equation}
S_B = k_{\rm B} \ln \frac{(2V)!}{(2V-2N)!}
\approx k_{\rm B} [2V \ln V -2N - 2(V-N) \ln (V-N) + 2N \ln 2]\, ,
\label{S_B}
\end{equation}
and hence the change in total entropy is
\begin{equation}
\Delta S \approx 2k_{\rm B} N \ln 2 \, .
\label{DeltaS}
\end{equation}

The above result shows two `unwanted' features.
First, being non-extensive, Boltzmann's statistical definition matches the thermodynamic entropy with difficulty.
Second, the total Boltzmann entropy increases after the partition is removed.  Since thermodynamics tells us that the state of the system (described through density, pressure, total number of particles, etc.) remains unchanged by the removal of the partition, this increase implies that the Boltzmann entropy cannot be an appropriate thermodynamic state variable.  Moreover, if one suddenly re-inserts the partition into the system, such that no work is done, the total Boltzmann entropy drops back down to its original value.  This again violates thermodynamics, where entropy can only be reduced through work and discharge of excessive heat in reversible processes, and must increase in
irreversible processes.

To resolve the Gibbs paradox and therefore define a statistical entropy that can be identified with the thermodynamic entropy, one replaces in the Boltzmann entropy formula the total number of configurations with the total number of `distinguishable' configurations.  Thus in classical cases (in which no two particles can have exactly the same coordinates in the phase space) one puts an extra $1/N!$ factor, while in quantum statistical mechanics one does the counting in a way that incorporate this indistinguishability inherently.

Statistical mechanics, or at least its classical version, is in fact a science concerning one's impression.  If the particles on the two sides of the partition are distinguishable, e.g., the particles on the left are red and the particles on the right are green, the total entropy should increase after the partition is removed.  This increase in entropy, and thus decrease in Helmholtz free energy, represents the least amount of work one has to perform in order to restore the system's `original' state, i.e., a situation that all the red particles are located on the left and all the green particles are on the right.  However, if one cannot tell the difference in color, no necessary action is needed to restore the system's original state, since the original state and the final state appear to be the same.

Even with such an understanding, one may still get puzzled if he considers other aspects of this story.  For example, one would note that
after the partition is removed, the accessible volume for each particle is twice as large as before.  Since it is encouraged by molecular chaos that each particle should access as much volume as possible, will this mechanism give rise to any driving force toward mixing, that is beyond our current description of thermodynamics?

There is yet another puzzle worthy of further studies.  If the partition is made of some fragile wall instead of an unbreakable one, one might find as time proceeds that the wall eventually breaks down due to collisions and the particles mix.  This phenomenon itself indicates the existence of a time arrow\cite{Lebowitz}, which implies an increase of some sort of entropy.  So where does this entropy increase come from?  Does it come merely from the dissociated wall molecules, or does it include any additional contribution from particle mixing?

Based on these speculations, we would like to ask whether thermodynamics, in particular {\it bulk} thermodynamics, gives correct prediction for such a system.  Is there any driving force towards particle mixing?  If the answer is yes, this driving force must imply some sort of entropy change after mixing.

\section{Analogy in a system with two colloidal particles}
For a colloidal particle in a salt-free aqueous solution, the size of its neutralizing counterion cloud is in the order of micrometers. Since there are about ten thousand neutralizing counterions surrounding each colloidal particle, the average distance among neighboring counterions is of several tens of nanometers, which is much larger than the Bjerrum length ($\approx 7\AA$ in aqueous solutions). Therefore the pairwise electrostatic interactions of counterions are rather weak, while the major force exerted on counterions are steric repulsions as well as the electrostatic confinement from all the other charges. Based on these facts we speculate that the volume-exclusive scenario as described in the previous section might be applied in the colloidal system.

Consider a system of two colloidal particles (named 'macroions' from now on, to avoid confusion with counterion particles), each dressed with a net charge of $-Ne$ on its surface and charge-compensated by $+N$ monovalent cations in the neighboring solution.  With no extra salt added, on average the $N$ counterions tend to reside in a vicinity volume $V$ of each macroion, and different counterion clouds do not overlap if the macroions sit far apart.

In this work we shall neglect the detailed electrostatic interaction between the macroions and their counterions, and represent the charge compensation by the previously mentioned simple scheme of lattice model.  In this scheme, the analogy of one macroion plus its $N$ counterions, is just $N$ particles in a box of volume $V$, as illustrated in Fig.~\ref{Fig1}.  The question is, if one considers two such macroions, is there any favored mixing entropy as suspected in the previous section, that drives the two macroions closer such that their counterions mix?

Consider a gedanken experiment in which the counterions of one macroion are painted 'red', while those of the other macroion are painted 'green'.  Assume that the counterions are otherwise identical.  The entropy of such a system will definitely increase when the macroions get closer and counterions mix.  But if this serves as a driving force that drags the two macroions closer, it would certainly apply as well for a system where the counterions are monochromatic, since the underlying microscopic mechanical rules (the Newtonian equations, for example) remain the same.  Such an increase of entropy associated with color mixing merely give us a hint about how much effort is needed to separate the mixture back into a red counterion cloud and a green one, while it requires more detailed argument to check whether there exists a driving force that drags the two counterion clouds closer.

%

In the next section we will show, with a detailed examination, that one can prove the existence an extra term $\delta S$ upon mixing, whether the counterions are all distinguishable or not.  Although this residual entropy is non-extensive, it does give rise to an important attraction in colloidal systems.

\section{Existence of an residual entropy and like-charge attraction}
\begin{figure}
  \includegraphics[width = 7cm]{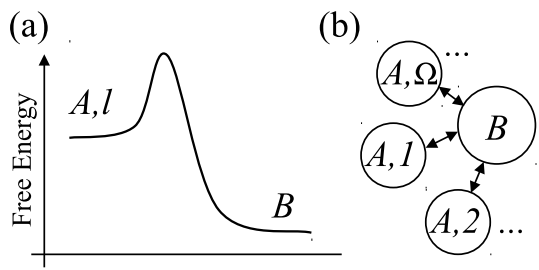}
  \caption{(a) Illustration of the free energy landscape that represents a transition between an arbitrary profile $(A,l)$ of the unmixed bisected state $A$ and the mixed state $B$.  The ``free energy'' is obtained from the Boltzmann entropy formula for distinguishable particles. (b) Transitions between state $B$ and all the profiles $(A,\{\})$ of the bisected state $A$.}
  \label{Fig2}
\end{figure}

Regardless of one's impression, the dynamical behavior of the physics-wise and chemistry-wise identical particles must be the same, whether they are distinguishable or not.  Keeping this in mind, we restrict our following discussion to the case in which all the particles are distinguishable.  To continue the argument in the previous sections, let us consider a state $A$ where the two macroions sit far apart such that the counterions could not transfer within the colloidal Brownian motion timescale, and a state $B$ where the two macroions are not far apart so that the counterions can diffuse to mix (see Fig.~\ref{Fig1}).  According to Boltzmann's entropy formula, the entropy of the mixed final state is higher, which amounts to a lower Helmholtz free energy, as shown in Fig.~\ref{Fig2}(a) (the internal energy is irrelevant here.)  For each profile $(A,l)$ of the initial unmixed state $A$, the entropy is $S_{A,l}$, and there exists a forward reaction rate constant $k_{f,l}$ towards the mixed state $B$, as well as a reverse reaction rate constant $k_{r,l}$.  The detailed-balance condition requires that
%
\begin{equation}
\frac{k_{f,l}}{k_{r,l}}= \exp \left( \frac{-\Delta F}{k_{\rm B}T} \right) = \exp \left( \frac{\Delta S}{k_{\rm B}} \right)
\approx 2^{2N} \, ,
\label{rate_ratio}
\end{equation}
where the approximation of $\Delta S$ is given by Eq.~\ref{DeltaS}.  This large ratio indicates the slim chance for the system to restore the initial profile $(A,l)$.  On the other hand, when the system is in state $B$, there also exist numerous demixing reactions towards other profiles $(A,l')$ (see Fig.~\ref{Fig2}(b)).  For the current case of two macroions, the total number of profiles is equal to the number of ways of bisecting $2N$ counterions.  Thus the total number of paths for these demixing reactions is $\Omega = (2N)!/(N!)^2 \approx 2^{2N}$.  While all these forward reactions share the same rate constant $k_f$, all the reverse reactions share another constant $k_r$.

Again according to the detailed-balance condition, the total forward reaction flux is equal to the total reverse reaction flux at thermal equilibrium, which gives
\begin{equation}
 P_{A,\{\}} \cdot k_f = P_B \cdot k_r \cdot \Omega \, ,
\label{detail_balance}
\end{equation}
where $P_{A,\{\}}$ means the total probability of the system's staying in the bisected profiles of state $A$, and $P_B$ is the probability of its staying in the mixed state $B$.  Eq.~\ref{rate_ratio} and Eq.~\ref{detail_balance} lead to the result that $P_{A,\{\}} \approx P_B$ at ``thermal equilibrium''.  Since $\exp (-F/k_{\rm B}T)$ tells about the probability up to some proportional constant, the Helmholtz free energy of $(A,\{\})$ and $B$ are approximately equal.

From the above argument one learns that at such ``thermal equilibrium'' the probability of observing bisected macroions is {\it approximately} identical to that of observing the macroions sitting close and counterions mixing up.

Intriguingly enough, one finds a more accurate estimate if one would care for a better off description of Stirling's approximation\cite{math_handbook}:
\begin{equation}
 \ln N! = \left( N+\frac12 \right)\ln (N+1) - (N+1) + \frac12 \ln (2\pi) + O\left( \frac1N \right) \, .
 \label{Stirling_fine}
\end{equation}
From the exact Boltzmann's entropy definition (equalities in Eqs.~\ref{Boltzmann}--\ref{S_B}), one obtains
\begin{equation}
 \frac{P_B}{P_{A,\{\}}} = \frac{k_f}{k_r \Omega}
 =  \frac{(2V)!}{(2N)!(2V-2N)!}  \left/ \left[ \frac{V!}{N!(V-N)!} \right]^2  \right.  \, ,
\end{equation}
which is exactly the same as the probability ratio obtained by treating all the counterions {\it indistinguishable}.  With the refined approximation in Eq.~\ref{Stirling_fine}, one derives
\begin{equation}
 \ln \frac{P_B}{P_{A,\{\}}} = \frac32 \ln N + \frac32 \ln \left(1-\frac{N}{V}\right) + \left[ \frac12 \ln (2\pi) - 1 - \frac32 \ln 2 \right] + O\left( \frac1N \right) \, .
\end{equation}
So there does exist a free energy difference between the mixed and bisected states, which means an increase in entropy:
\begin{equation}
 \delta S  \approx k_{\rm B} \left\{ \frac32 \ln N + \frac32 \ln \left(1-\frac{N}{V}\right) + \left[ \frac12 \ln (2\pi) - 1 - \frac32 \ln 2 \right] \right\} \, .
\label{residual_S}
\end{equation}

A remarkable feature of $\delta S$ in Eq.~\ref{residual_S} lies in the fact that it is non-extensive.  Thus the contribution of $\delta S$ becomes irrelevant when one considers bulk thermodynamics.  However, as we shall find in the next paragraph, for a system such as a pair of macroions plus their counterions, where the number of counterions surrounding each macroion is in the order of $10^4$, the contribution of $\delta S$ is approximately $12 k_{\rm B}$ upon mixing.  In other words, the free energy is lowered by $\sim 12k_{\rm B}T$ when the macroions approach each other.  This is large compared with thermal fluctuations of the colloidal particles, and therefore the residual mixing entropy can be a crucial candidate among the sources of the effective like-charge attraction.

To estimate the magnitude of this residual mixing entropy, we first approximate the thickness of the counterion cloud by the Debye screening length:
\begin{equation}
  \lambda \approx \frac{1}{\kappa} = \frac{1}{\sqrt{4\pi l_B\cdot 2\rho_w}}\, ,
\end{equation}
where $\rho_w$ is the density of dissociated water ions, as one has $\rho_w \approx 60 \mu m^{-3}$.  As we estimate the size of a counterion by the Bjerrum length, which indicates the electrostatic strength, the corresponding volume of occupation is approximately $(l_B/2)^3\cdot 4\pi/3$.  Thus the effective packing fraction of counterions is
\begin{equation}
  \rho_c = \frac{N}{V} = \left.  N\cdot \frac43 \pi \left(\frac{l_B}{2}\right)^3 \middle/ \left\{\frac43 \pi \left[ (R+\lambda)^3 - R^3\right]\right\} \right.\, .
  \label{NV}
\end{equation}

\begin{figure}
  \includegraphics[width = 7cm]{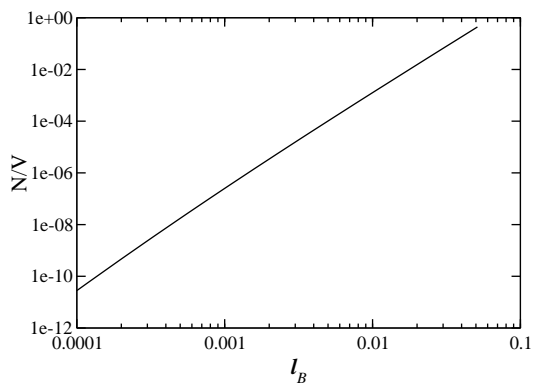}
  \caption{Effective packing fraction of counterions versus the Bjerrum length. We use the approximation that each counterion occupies a volume $(l_B/2)^3\cdot 4\pi/3$.}
  \label{Fig3}
\end{figure}

\begin{figure}
  \includegraphics[width = 7cm]{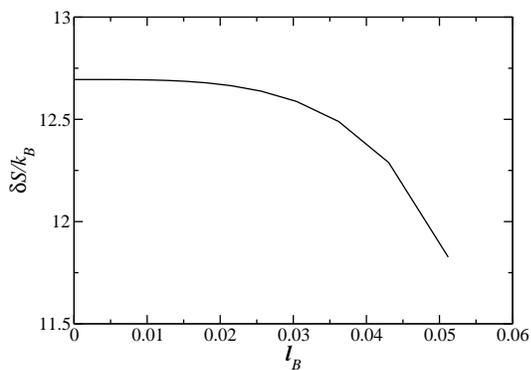}
  \caption{The residual mixing entropy $\delta S$ versus the Bjerrum length.}
  \label{Fig4}
\end{figure}

Fig.~\ref{Fig3} shows the dependence of the counterion volume fraction $\rho_c$  on the Bjerrum strength.  Note that as the volume fraction approaches one, the electrostatic interaction among neighboring counterions becomes relevant and our simple model will fail at such regime.  Nevertheless, as in a normal aqueous solution, where $l_B \approx 7\AA$, $\rho_c = 6.4\times 10^{-8}$, the distance between neighboring counterions is quite large such that the ideal-gas approximation can be applied in this regime.  Meanwhile, our result about the residual entropy is presented in Fig.~\ref{Fig4}.  We find that $\delta S \approx 12.7 k_{\rm{B}}$ at low Bjerrum length.  The magnitude of this residual entropy stays constant throughout the observed regime.

The analogy we have made, between the system of two macroions with counterions and the scenario of the Gibbs paradox, is based on the assumption that the counterions cannot transfer from one cloud to the other when the macroions are far apart, as if there exists a partition in between.  As the macroions get closer, the counterions mix, as if the partition is removed.  Our result tells that the residual entropy is exactly the contribution from the number fluctuation of counterions upon mixing.  Indeed, one could argue that such number fluctuations would possibly lead to oppositely-charged colloidal particles that attract each other via simple Coulomb interaction.  However, we should keep in mind that for this Coulomb interaction to be comparable to the order of thermal energy, with the colloidal particle separation of a few microns, one needs to examine a scenario of non-neutral particles charged as much as $\sim \pm 100e$.  Furthermore, the total electrostatic energy in such a case should also take into account the ionization energy as well as energy of affiliation for each counterion jumping from one cloud to another, as the net electrostatic energy will be less significant.  Also it has been known that the induced dipole-dipole interaction at such large colloidal separations is weaker than the repulsive DLVO potential\cite{Levin02}.  Our mechanism of residual mixing entropy may provide a source for the like-charge attraction between colloidal particles, based on the speculation that its magnitude is comparable to the thermal energy.


\section{Conclusion}
Through a re-examination of the Gibbs paradox, we deduce the existence of a non-extensive residual entropy $\delta S$ upon particle mixing, even for cases where particles are indistinguishable.  Comparing the mixing of particles in Gibbs paradox to the mixing of counterions in a colloidal system, we speculate that this residual entropy leads to an effective attraction among colloidal particles.  This attraction, contributed from the residual entropy upon mixing, arises from the number fluctuation of counterions when the colloidal particles approach each other.


Throughout our discussion, the details of electrostatic interaction are neglected except for the overall charge neutrality between each colloidal particle and its counterion cloud.  Whether these details can enhance the residual entropy effect is not known yet, as this will be further studied in our future work.  Moreover, our arguments in the previous sections are based on the speculation that the case where counterion clouds are far apart can be compared to a bisected state, due to the consideration of overall charge neutrality about each colloidal particle.  In fact, even if the number of counterions can still fluctuate, our proposed mechanism of like-charge attraction may be observed as a nonequilibrium process.  This is because it takes a relatively large time for counterions to diffuse over clouds, and the mixing of counterions between clouds may be incomplete over the observed time scale.  To clarify these effects, we suggest that a thorough investigation via a Poisson-Boltzmann approach may help resolve this problem with better accounts.
Concerning the original Gibbs paradox, we conclude that the Boltzmann entropy serves as a good definition of the `statistical' entropy, as it faithfully tells about the probability distribution and the direction of the time arrow, and leaves no paradox in the end of the argument.  On the other hand, due to its non-extensiveness and violation of the second law, it does not serve as an appropriate definition of the bulk `thermodynamic' entropy,  despite that it still satisfies the variational statement of the second law\cite{Chandler}, which states that the entropy of a system is smaller if there exists internal constraints.

\ack
This work was supported by the National Science Council of the Republic of China under Grant No. NSC-98-2112-M-008-009, and No. 99-2112-M-003-012-MY2. CLL would like to thank for support from NCTS under focus group Life and Complexity.


\end{document}